\documentclass[sigconf]{acmart}

\AtBeginDocument{%
  }

\setcopyright{acmlicensed}
\copyrightyear{2027}
\acmYear{2027}
\setcopyright{cc}
\setcctype{by}
\acmConference[WSDM '27]{Proceedings of the Twentieth ACM International Conference on Web Search and Data Mining}{February 15--19, 2027}{Hong Kong, Hong Kong}
\acmBooktitle{Proceedings of the Twentieth ACM International Conference on Web Search and Data Mining (WSDM '27), February 15--19, 2027, Hong Kong, Hong Kong}
\acmDOI{10.1145/3862322.3863809}
\acmISBN{979-8-4007-3117-4/2027/02}

\usepackage{multicol}
\usepackage{enumitem}
\usepackage{multirow}
\usepackage{booktabs}
\usepackage{array}
\usepackage{subfigure}
\usepackage{float}
\usepackage{stfloats}
\usepackage{balance}
\usepackage{marvosym}

\begin{document}

\title{Beyond Post-hoc Rationales: Improving Reasoning SFT for LLM-based Recommendation}


\author{Yufei Ye}
\affiliation{%
  \institution{University of Science and Technology of China}
  \city{Hefei}
  \country{China}}
\email{aboluo2003@mail.ustc.edu.cn}

\author{Jiaxin Deng}
\affiliation{%
  \institution{Kuaishou Technology}
  \city{Beijing}
  \country{China}}
\email{dengjiaxin03@kuaishou.com}

\author{Xingmei Wang}
\affiliation{%
  \institution{Kuaishou Technology}
  \city{Beijing}
  \country{China}}
\email{wangxingmei@kuaishou.com}

\author{Yingzhi He}
\affiliation{%
  \institution{Kuaishou Technology}
  \city{Beijing}
  \country{China}}
\email{heyingzhi@kuaishou.com}

\author{Jian Liang}
\affiliation{%
  \institution{Kuaishou Technology}
  \city{Beijing}
  \country{China}}
\email{liangjian03@kuaishou.com}

\author{Ruiming Tang}
\affiliation{%
  \institution{Kuaishou Technology}
  \city{Beijing}
  \country{China}}
\email{tangruiming@kuaishou.com}

\author{Cheng Luo}
\affiliation{%
  \institution{Kuaishou Technology}
  \city{Beijing}
  \country{China}}
\email{luocheng10@kuaishou.com}

\author{Hao Wang}
\affiliation{%
  \institution{University of Science and Technology of China}
  \city{Hefei}
  \country{China}}
\email{wanghao3@ustc.edu.cn}

\author{Defu Lian}
\affiliation{%
  \institution{University of Science and Technology of China}
  \city{Hefei}
  \country{China}}
\email{liandefu@ustc.edu.cn}
\authornote{Corresponding author.}

\author{Enhong Chen}
\affiliation{%
  \institution{University of Science and Technology of China}
  \city{Hefei}
  \country{China}}
\email{cheneh@ustc.edu.cn}

\renewcommand{\shortauthors}{Yufei Ye et al.}

\begin{abstract}
    Large language models equipped with Semantic IDs have enabled LLM-based generative recommendation by formulating next-item prediction as autoregressive token generation. Recent systems have begun to introduce CoT-style rationales into training, yet at the SFT stage their reasoning mode often lags behind direct prediction. We identify three causes: target-exposed CoT synthesis produces post-hoc rationales that are hard for the student to reproduce at inference; text-level CoT filters poorly reflect recommendation quality; and standard SFT dilutes supervision on the final target SID with long rationale tokens. We propose R$^2$-SFT, a reasoning-enhanced SFT framework for recommendation foundation models. R$^2$-SFT first synthesizes target-blind rationales from user history only, then uses a warm-up recommender as a SID-perplexity scorer to select rationales under a criterion aligned with the recommendation objective, and finally adds a target-aware loss that strengthens supervision on the recommended SID. Extensive experiments on RecIF-Bench demonstrate that R$^2$-SFT consistently enables reasoning-mode recommendation to surpass its non-reasoning counterpart at the SFT stage, and improves Pass@32 by 69\% on average over the OpenOneRec-SFT reasoning baseline. Code is available at \url{https://github.com/aboluo2003/R-2-SFT}.
\end{abstract}




\maketitle

\section{Introduction}
\label{sec:introduction}

Large Language Models (LLMs) have demonstrated strong reasoning ability through chain-of-thought (CoT)~\cite{wei2022chain, guo2025deepseek, kojima2022large}, which improves performance on a variety of tasks by eliciting a step-by-step inference process before the final answer. This naturally motivates recent foundation models for recommendation to adopt the same paradigm and elicit an explicit thought process before predicting the next item~\cite{zheng2026recgpt, liu2025onerec, li2026gr2}. However, equipping a recommender with an explicit reasoning step does not automatically translate into a performance gain. For OpenOneRec~\cite{zhou2025openonerec}, the reasoning mode still trails the non-reasoning mode by about 20\% after reinforcement learning (RL), and the gap is even larger at the supervised fine-tuning (SFT) stage. Although OneReason~\cite{team2026onereason} adopts a specialize-then-unify reinforcement learning recipe that enables the reasoning mode to surpass its non-reasoning counterpart, this approach demands substantial training compute. \citet{zhang2026thinking} study the same problem and propose a training-free remedy, but their method incurs additional inference cost, further burdening the already expensive reasoning mode. We therefore ask whether the SFT stage itself can be improved to give a more fundamental solution.

We approach this question from both the data side and the training side, and identify three main reasons why prior SFT recipes such as OneReason~\cite{team2026onereason} and OpenOneRec~\cite{zhou2025openonerec} fail to make the reasoning mode outperform the non-reasoning mode at the SFT stage. First, the CoT synthesis process introduces bias: a common practice exposes the target item to the teacher model when synthesizing the rationale~\cite{team2026onereason, he2026reasoning, liang2026gr2}, so the teacher is asked to explain an already-known answer rather than to infer the next item from history, yielding post-hoc rationales~\cite{peng2026measuring} that are difficult for the student to generalize. Second, there is no effective way to filter CoT for generative recall. Directly aligning with recommendation metrics such as NDCG requires beam search, which is slow and has sparse hits, making it too costly to produce enough CoT at scale; existing alternatives therefore fall back on text-level signals such as text similarity~\cite{wang2026mllmrec} or text perplexity~\cite{li2026taiji}. However, there is a gap between text and Semantic-ID (SID) representations, and the LLM judge used for evaluation is not specialized for recommendation, so text-level filtering does not truly reflect whether a rationale helps recommendation. Third, the supervision objective at training time is misaligned with the recommendation task: during SFT the supervision signal is diluted by the CoT text, so the model only learns to generate fluent rationales rather than to recommend the target item correctly while thinking.

To address the above issues, we propose \textbf{R$^2$-SFT}, a reasoning-enhanced SFT framework. R$^2$-SFT avoids exposing the target item when synthesizing CoT, so that the rationale reflects a genuine inference from history. To filter the synthesized CoT, R$^2$-SFT warm-starts the base model on a small portion of CoT data to adapt it to the reasoning format, then scores each rationale by the perplexity of the target SID as a recommendation-aligned filtering criterion. Finally, R$^2$-SFT introduces an additional target loss during training to strengthen the supervision on the target SID and mitigate the dilution of the recommendation signal by the CoT text.

Our contributions are as follows.
\begin{itemize}[leftmargin=*,align=left]
    \item We study and characterize the phenomenon that the reasoning mode hurts LLM-based recommenders at the SFT stage, identify its root causes from both the data and training sides, and accordingly propose the R$^2$-SFT framework to solve it.
    \item On the data side, we propose target-blind CoT synthesis to prevent the teacher from producing post-hoc rationales, and SID-PPL rejection sampling that filters CoT efficiently while being aligned with the recommendation objective.
    \item On the training side, we introduce an additional target loss in the SFT stage to address the dilution of the recommendation signal by the CoT text.
    \item We conduct extensive experiments on public industrial data spanning three domains, and the results show that R$^2$-SFT consistently enables the reasoning mode to outperform the non-reasoning mode at the SFT stage.
\end{itemize}

\section{Related Work}
\label{sec:related_work}

\subsection{LLM-Based Recommendation}
\label{subsec:llm_based_rec}

Large Language Models (LLMs) encode rich world knowledge and exhibit robust semantic understanding, which can alleviate long-standing challenges in traditional recommender systems, such as the cold-start problem and data sparsity \cite{wu2024survey, zhao2024recommender}. Consequently, leveraging LLMs for recommendation has become an active research direction.

Early studies \cite{liu2023chatgpt, gao2023chat} attempted to elicit recommendation capabilities from LLMs via prompt engineering. While these approaches achieved certain gains by exploiting the inherent knowledge of LLMs, their performance remained suboptimal due to the lack of domain-specific knowledge in recommendation. To address this, TALLRec~\cite{bao2023tallrec} and LLaRA~\cite{liao2024llara} incorporate recommendation-specific knowledge into LLMs through supervised fine-tuning.

Subsequently, generative recommendation \cite{rajput2023recommender, deng2025onerec, zhou2025onerec} based on Semantic IDs (SIDs) \cite{hou2023learning, ge2013optimized, mentzer2024finite} has gained momentum. Since SIDs preserve the original vocabulary of LLMs by avoiding the introduction of an excessive number of new tokens, while also enabling efficient candidate item generation, they have been widely adopted in LLM-based recommendation. LC-Rec~\cite{zheng2024adapting} fine-tunes LLaMA~\cite{touvron2023llama} to enable LLMs to comprehend SID tokens while simultaneously learning knowledge from the recommendation domain. PLUM~\cite{he2026plum}, on the other hand, injects both recommendation domain knowledge and an understanding of SID tokens into LLMs through continued pre-training. OpenOneRec~\cite{zhou2025openonerec} retains general-domain capabilities while incorporating recommendation-specific knowledge, thereby endowing the model with novel functionalities such as explainable recommendation and interactive recommendation.

\subsection{Reasoning-Enhanced Recommendation}
\label{subsec:reasoning_rec}

Inspired by large reasoning models such as o1~\cite{jaech2024openai} and DeepSeek-R1~\cite{guo2025deepseek}, the recommendation community has begun to leverage reasoning to push the performance ceiling of recommender systems. Reasoning in recommendation can be roughly categorized into implicit reasoning and explicit reasoning.

Implicit reasoning~\cite{zhang2026reinforced, lin2026bringing, tang2026think} iteratively refines latent states in a continuous space to raise the upper bound of recommendation performance. Although effective, its reasoning process remains a black box and is difficult to interpret, which motivates the explicit, language-grounded reasoning studied in this work.

Explicit reasoning employs natural language as the medium and adopts chain-of-thought (CoT) style inference. However, direct CoT data specific to the recommendation domain are scarce. Existing work often resorts to synthesizing such data using teacher models~\cite{ho2023large, team2026onereason, liang2026gr2, fang2025reason4rec}. The data acquisition process generally consists of two steps: synthesis and filtering. For synthesis, a considerable line of work~\cite{yu2026thinkrec, liu2025onerec, team2026onereason, zhou2025openonerec, he2026reasoning} generates CoT by exposing the target item to the teacher model. For filtering, in ranking scenarios, the availability of a small candidate set or explicit ratings enables direct filtering based on ranking/rating outcomes~\cite{fang2025reason4rec, liang2026gr2}. In recall scenarios, however, there is no direct evaluation criterion; existing approaches commonly adopt text-based methods~\cite{wang2026mllmrec, li2026taiji}. For instance, MLLMRec-R1~\cite{wang2026mllmrec} employs text similarity, while Taiji~\cite{li2026taiji} uses text perplexity. Nevertheless, two issues arise: (1) synthesizing CoT by exposing the target may lead to post-hoc explanations that do not generalize well; (2) text similarity or perplexity cannot effectively reflect recommendation quality.

On the training side, most explicit-reasoning works rely on reinforcement learning to unlock the reasoning potential~\cite{team2026onereason, he2026reasoning, you2026r, li2026taiji, liang2026gr2}. Exploration at the supervised fine-tuning (SFT) stage remains limited; within a single model, existing efforts that incorporate reasoning during SFT often find the thinking mode trailing its non-thinking counterpart~\cite{team2026onereason, zhou2025openonerec, zhang2026thinking}.

\section{Preliminaries} 

\subsection{Semantic ID for LLM-based Recommendation}

Representing items in LLM-based recommendation is non-trivial: text-based representations suffer from excessively verbose inputs and a lack of collaborative signals, whereas traditional numerical IDs introduce an overly large vocabulary that disrupts the LLM's pre-trained token distribution \cite{hua2023index, zheng2024adapting,liao2024llara}. To resolve this, Semantic IDs (SIDs) serve as a compact alternative, mapping each item into a short sequence of discrete tokens derived from its semantic content, thus maintaining a manageable vocabulary while preserving rich item information.

Formally, given an item $i \in \mathcal{I}$, we first obtain its continuous semantic embedding $\mathbf{x}_i \in \mathbb{R}^d$ using a pre-trained encoder. A residual quantization module (e.g., RQ-VAE~\cite{lee2022autoregressive}) is then employed to discretize this embedding. The module consists of $m$ separate codebooks $\{\mathcal{C}_1, \mathcal{C}_2, \dots, \mathcal{C}_m\}$, and the Semantic ID of item $i$ is defined as a tuple of codewords:
\begin{equation}
    \mathbf{z}_i = (c_{i,1}, c_{i,2}, \dots, c_{i,m}), \quad \text{where } c_{i,j} \in \mathcal{C}_j.
\end{equation}
Here, $m$ denotes the length of the Semantic ID. Following common practices in generative recommendation, we set $m=3$ in this work.

Within this framework, the LLM acts as an autoregressive sequence predictor. Given a user's historical interactions:
\begin{equation}
    \mathcal{H}_u = \langle i_1, i_2, \dots, i_t \rangle,
\end{equation}
the sequence is first translated into a sequence of Semantic IDs, denoted as $\mathbf{Z}_u = \langle \mathbf{z}_{i_1}, \mathbf{z}_{i_2}, \dots, \mathbf{z}_{i_t} \rangle$. The LLM, parameterized by $\theta$, processes this sequence and autoregressively generates the next item's Semantic ID. The predictive probability is factorized over the $m$ codewords:
\begin{equation}
    P_\theta(\mathbf{z}_{i_{t+1}} \mid \mathbf{Z}_u) = \prod_{j=1}^{m} P_\theta\left(c_{i_{t+1}, j} \mid \mathbf{Z}_u, c_{i_{t+1}, 1}, \dots, c_{i_{t+1}, j-1}\right).
\end{equation}
This formulation aligns the recommendation objective with the LLM's native autoregressive paradigm, enabling it to leverage powerful sequence modeling capabilities to capture user preferences and generate recommendations directly within a compact semantic token space, rather than introducing an intractable number of unstructured item tokens.

\subsection{Reasoning-Enhanced Recommendation}

The core of reasoning-enhanced recommendation lies in eliciting a deliberate inference process prior to the final prediction. Formally, given a user's historical interaction sequence $\mathcal{H}_u = \langle i_1, i_2, \dots, i_t \rangle$, we first translate it into a Semantic ID sequence $\mathbf{Z}_u$. The model's objective is then to generate reasoning tokens $\tau$, followed by predicting the next Semantic ID, $\mathbf{z}_{i_{t+1}}$.

This generative process can be formulated as marginalizing over the reasoning paths:
\begin{equation}
    P_\theta(\mathbf{z}_{i_{t+1}} \mid \mathbf{Z}_u) = \sum_{\tau} P_\theta(\tau \mid \mathbf{Z}_u) \cdot P_\theta(\mathbf{z}_{i_{t+1}} \mid \mathbf{Z}_u, \tau).
\end{equation}
The reasoning tokens $\tau$ serve as an intermediate thought process, summarizing user intent, filtering potential candidates, or inferring preferences, thereby guiding the model toward a more accurate final recommendation.

\begin{figure*}[htb]
    \centering
    \includegraphics[width=\textwidth]{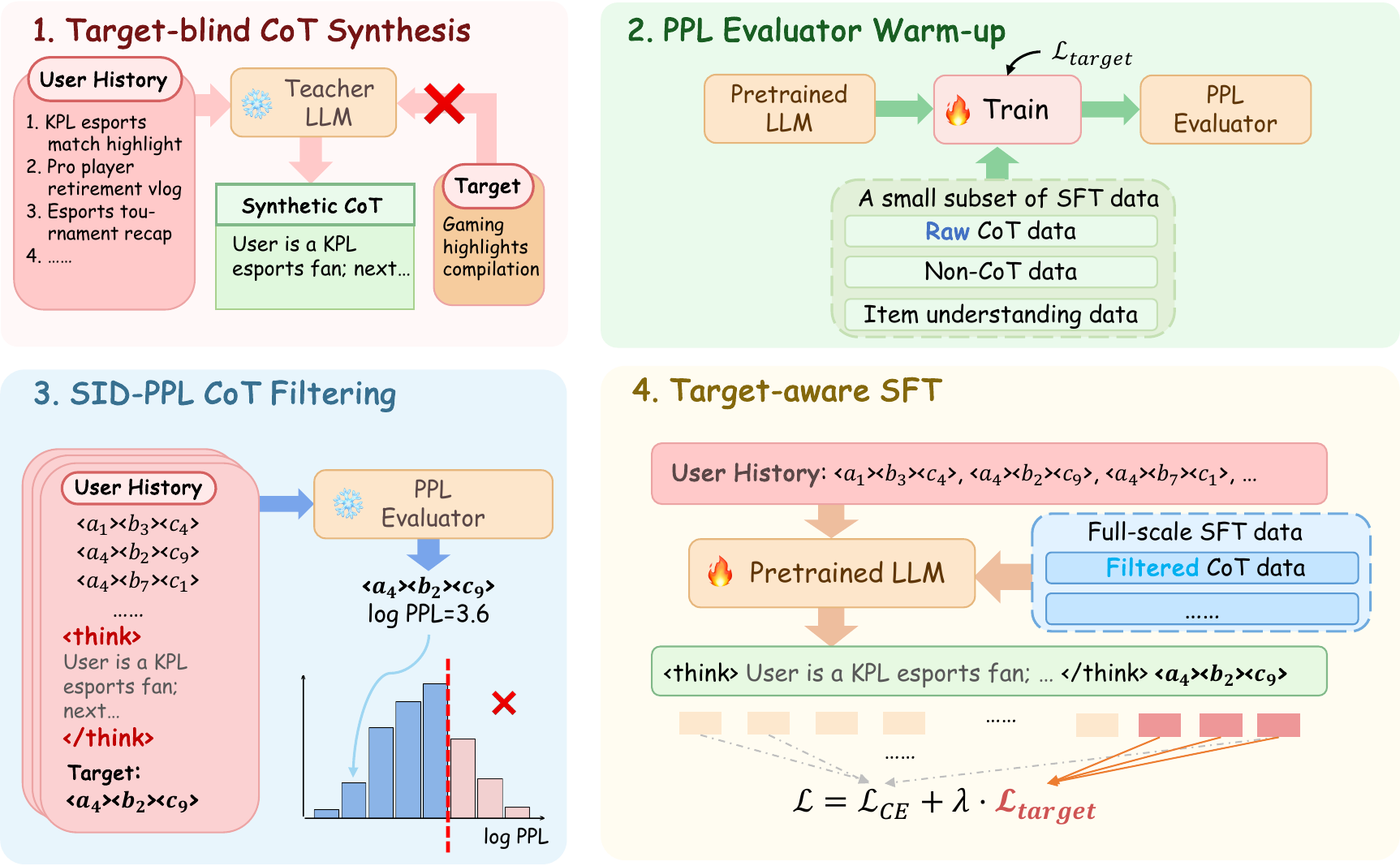}
    \caption{Overview of R$^2$-SFT. Stage~1: a teacher LLM synthesizes CoT from the user history without access to the target item. Stage~2: a small subset of the SFT data warm-starts the base model into a PPL scorer for thinking-format sequences. Stage~3: the scorer computes the target-SID log PPL for each candidate rationale, and low-PPL rationales are kept. Stage~4: the filtered CoT, combined with item-understanding and non-CoT data, is used for target-aware SFT with an added loss on the final item SID.}

    \label{fig:r2sft}
\end{figure*}

\section{Methodology}
\label{sec:methodology}

In this section, we present \textbf{R$^2$-SFT}, our framework for reasoning-enhanced LLM-based recommendation, as illustrated in Figure~\ref{fig:r2sft}. Our method is built on the publicly released OpenOneRec-8B-Pretrain\footnote{https://huggingface.co/OpenOneRec/OneRec-8B-pretrain}, which has already acquired basic recommendation ability without explicit reasoning and can understand and generate Semantic IDs (SIDs). Starting from this checkpoint, R$^2$-SFT improves reasoning-style recommendation through four stages: (i) target-blind CoT synthesis, which constructs rationales without exposing the target to the teacher; (ii) PPL scorer warm-up, which adapts the base model to thinking-format sequences so it can score rationales reliably; (iii) SID-PPL CoT filtering, which keeps rationales that genuinely help target SID prediction; and (iv) target-aware SFT, which strengthens the supervision on the final recommended item. 

\subsection{SFT Data Composition}
\label{subsec:sft_data}

R$^2$-SFT uses a mixed supervised fine-tuning corpus consisting of three types of data. The purpose of this mixture is to improve reasoning-based recommendation while avoiding degradation of the SID grounding and direct recommendation ability obtained from the pre-training stage.

\paragraph{Item understanding data}
The first part contains item understanding tasks that align SIDs with natural-language item semantics. Specifically, we include both \texttt{SID2Caption} and \texttt{Caption2SID} tasks. In \texttt{SID2Caption}, the model is asked to generate the textual caption of an item given its SID. In \texttt{Caption2SID}, the model predicts the corresponding SID given the item caption. These two directions help maintain the bidirectional alignment between natural-language descriptions and discrete SID tokens.

\paragraph{Direct recommendation data without CoT}
The second part consists of standard recommendation examples without chain-of-thought reasoning. Given the user's historical interaction sequence $\mathcal{H}_u$ mapped to the Semantic ID sequence $\mathbf{Z}_u$, the model is trained to directly predict the next item SID, denoted as $\mathbf{z}_{i_{t+1}}$. This component serves to preserve the fundamental recommendation capabilities acquired during the pre-training stage.

\paragraph{Reasoning-based recommendation data with CoT}
The third part incorporates recommendation examples augmented with reasoning traces. Since the base pre-trained model lacks the innate ability to perform explicit reasoning in the recommendation context, this subset is designed to equip the model with thinking capabilities. Similar to the non-CoT recommendation data, the input consists of the SID sequence $\mathbf{Z}_u$. However, the target output extends the direct prediction by including an intermediate rationale $\tau$ in the natural language space, followed by the ground-truth target item SID $\mathbf{z}_{i_{t+1}}$. Consequently, the model learns to first perform explicit reasoning based on the user history and then predict the target SID.

\subsection{Target-blind CoT Synthesis}
\label{subsec:target_blind_cot}

The pre-trained model has strong recommendation ability but has never been trained on reasoning data in the recommendation domain, so it cannot perform reasoning over SID sequences before prediction. To equip the model with reasoning ability on the recommendation task, we use an external teacher LLM $P_{\phi}$ to synthesize chain-of-thought rationales in the natural-language space, and let the student learn to map the SID history to the rationale and then to the target SID. In this process, the model input is replaced with the SID sequence used by the recommender, while the CoT supervision is provided by the teacher-synthesized natural-language rationale.

Concretely, given a user history $\mathcal{H}_u = \langle i_1, \dots, i_t \rangle$ with the corresponding SID sequence $\mathbf{Z}_u$, we convert each interacted item $i_k$ into its textual caption $c(i_k)$ and feed the caption sequence to the teacher. The teacher is prompted to analyze the user's behavior, infer the user's preferences, and predict the type and attributes of the next item, without access to the target item or any item identifier. The rationale is sampled as:
\begin{equation}
    \tau \sim P_{\phi}\!\left(\tau \mid c(i_1), \dots, c(i_t),\, \Omega \right),
\end{equation}
where $\Omega$ denotes the instructions. The generated rationale $\tau$ is a natural-language paragraph that describes the inferred preferences.

A key design choice is that the teacher must not see the real target $\mathbf{z}_{i_{t+1}}$. Several recent works~\cite{team2026onereason, he2026reasoning, liang2026gr2} synthesize CoT with the target exposed to the teacher, so the teacher can write a rationale that fits the known answer. Such post-hoc rationales raise a faithfulness concern similar to prior observations on unfaithful CoT explanations~\cite{turpin2023language}: they may teach the model to justify an already-decided prediction rather than to infer the next item from the history, leading to poor generalization to unseen targets (Section~\ref{subsec:generalization}). To avoid this, we adopt a target-blind protocol: the teacher is given only the history captions, so that the rationale reflects a genuine inference from the history rather than a post-hoc explanation of $\mathbf{z}_{i_{t+1}}$.

However, because the rationale is produced without seeing the target, it is not guaranteed to be consistent with the ground-truth $\mathbf{z}_{i_{t+1}}$: a plausible rationale may still predict a wrong direction. We therefore filter the synthesized CoT in Section~\ref{subsec:ppl_filtering} using a SID-level criterion that measures whether $\tau$ actually helps predict $\mathbf{z}_{i_{t+1}}$.

\subsection{SID-PPL-based CoT Filtering}
\label{subsec:sid_ppl_filtering}

Target-blind CoT synthesis avoids target leakage, but the synthesized rationales still vary in quality: a rationale may be fluent yet irrelevant to the final recommendation. Filtering is therefore needed to keep rationales that genuinely help prediction. Existing filtering methods are mainly text-based~\cite{wang2026mllmrec, li2026taiji}, relying on text similarity, text perplexity, or LLM-Judge scores. However, there is a gap between the text space and the SID space, and a general LLM judge does not understand recommendation, so text-level quality does not reflect whether a rationale helps predict the target SID. An ideal alternative would be to directly align filtering with recommendation metrics such as NDCG, but this requires beam search over the SID vocabulary for every candidate, which is inefficient and yields sparse hits, making large-scale CoT production costly.

We instead filter CoT by SID perplexity: for each candidate rationale, we measure how well it helps the model predict the target SID in a single forward pass. SID-PPL is directly aligned with the recommendation objective---it evaluates exactly the quantity the model is trained to predict---and is efficient to compute at scale. The filtering pipeline has two steps: we first warm up the pre-trained model into a PPL scorer adapted to the thinking-format input (Section~\ref{subsec:ppl_warmup}), and then use this scorer to compute the target-SID PPL of each candidate and reject low-quality rationales (Section~\ref{subsec:ppl_filtering}).

\subsubsection{PPL Scorer Warm-up}
\label{subsec:ppl_warmup}

The pre-trained model has never seen the thinking-format input that combines a SID history, a natural-language rationale, and a SID target, so directly using it to score $(\mathbf{Z}_u, \tau, \mathbf{z}_{i_{t+1}})$ triples would yield unreliable perplexity estimates for the target SID under this unfamiliar context. We therefore adjust the model with a small amount of data before using it as a scorer. Since the pre-trained model has already learned rich item knowledge and recommendation ability, only a small amount of data is needed to help it adapt to the thinking-format recommendation input (i.e., reasoning-augmented input). Concretely, we conduct a short warm-up SFT on about $6\%$ of the full training data, using the same mixed corpus as in Section~\ref{subsec:sft_data} with unfiltered CoT, and train it with the target-aware SFT objective introduced in Section~\ref{subsec:target_loss}. The resulting warm-up checkpoint is used as the scorer $P_\theta$ in the filtering step.

\subsubsection{SID-PPL CoT Filtering}
\label{subsec:ppl_filtering}

For each candidate reasoning trace $\tau$, we compute the perplexity of the target SID $\mathbf{z}_{i_{t+1}} = (c_1, \dots, c_m)$ conditioned on the SID history and the rationale:
\begin{equation}
    \mathrm{PPL}_{\mathrm{SID}}(\mathbf{z}_{i_{t+1}} \mid \mathbf{Z}_u, \tau)
    = \exp\left(-\frac{1}{m}\sum_{j=1}^{m}
    \log P_{\theta}(c_j \mid \mathbf{Z}_u, \tau, c_{<j})\right).
\end{equation}
A lower SID-PPL indicates that $\tau$ provides more useful evidence for predicting the target SID, so we keep low-PPL rationales and reject high-PPL ones. In practice, to simplify computation, we directly use $\log \mathrm{PPL}_{\mathrm{SID}}(\mathbf{z}_{i_{t+1}} \mid \mathbf{Z}_u, \tau) = -\frac{1}{m}\sum_{j=1}^{m}\log P_{\theta}(c_j \mid \mathbf{Z}_u, \tau, c_{<j})$ for filtering; the two are monotonically related and lead to identical selection. We adopt an empirical retention ratio and keep a fixed percentile of candidates with the lowest SID-PPL as the filtered CoT set.

\subsection{Target-aware SFT}
\label{subsec:target_loss}

For CoT recommendation examples, the rationale $\tau$ is much longer than the target SID $\mathbf{z}_{i_{t+1}}$. If the model is optimized only with the standard next-token loss over the entire output sequence, the supervision on the target SID is diluted by the long rationale, and the model may learn to produce fluent rationales without sufficiently improving the final recommendation accuracy. To address this, R$^2$-SFT introduces an additional \emph{target loss} that strengthens the supervision on the target SID. The target loss is applied only to CoT recommendation examples; the item-understanding and direct-recommendation data from Section~\ref{subsec:sft_data} are trained with the standard cross-entropy loss alone.

Concretely, let the output sequence of a CoT recommendation example be $o = [\tau; \mathbf{z}_{i_{t+1}}] = (o_1, \ldots, o_N)$ of length $N$, where $\mathbf{z}_{i_{t+1}} = (c_1, \dots, c_m)$ occupies the last $m$ tokens. The standard cross-entropy loss is the mean over all output tokens, and the target loss is the mean over only the target SID tokens:
\begin{equation}
    \mathcal{L}_{\mathrm{CE}}
    = - \frac{1}{N}\sum_{k=1}^{N}
    \log P_{\theta}(o_k \mid \mathbf{Z}_u, o_{<k}),
\end{equation}
\begin{equation}
    \mathcal{L}_{\mathrm{target}}
    = - \frac{1}{m}\sum_{j=1}^{m}
    \log P_{\theta}(c_j \mid \mathbf{Z}_u, \tau, c_{<j}).
\end{equation}
The training objective for CoT recommendation examples is
\begin{equation}
    \mathcal{L}= \mathcal{L}_{\mathrm{CE}} + \lambda \mathcal{L}_{\mathrm{target}},
\end{equation}
where $\lambda$ controls the strength of the target supervision. For item-understanding and direct-recommendation examples, the objective is the standard cross-entropy loss $\mathcal{L}_{\mathrm{CE}}$ without the target loss. The final training corpus combines all three data types from Section~\ref{subsec:sft_data}, where the CoT recommendation part uses the filtered rationales from Section~\ref{subsec:ppl_filtering}.

\section{Experiments}
\label{sec:experiments}

\subsection{Experimental Setup}
\label{subsec:setup}

\subsubsection{Datasets and metrics.}
Following prior work \cite{zhou2025openonerec}, we conduct experiments on RecIF-Bench\footnote{https://huggingface.co/datasets/OpenOneRec/OpenOneRec-RecIF}, a holistic recommendation instruction-following benchmark that spans three heterogeneous industrial domains: Short Video, Ad, and Product. RecIF-Bench aggregates nearly 100M interactions from more than 100K users, and uses a strict user-based split to prevent data leakage. Dataset statistics are summarized in Table~\ref{tab:dataset_stats}. Following the OpenOneRec evaluation protocol, we report Pass@32 and Recall@32 on the next-item prediction task, abbreviated as P@32 and R@32 in compact tables and figures. Pass@32 measures whether the ground-truth item appears among the top-32 generated candidates, while Recall@32 measures the proportion of ground-truth items retrieved. The OpenOneRec variants support both a thinking mode and a non-thinking mode, and we evaluate each variant in both modes; the remaining baselines operate in a single mode only. For brevity, we refer to the Short Video domain as Video in tables and figures.

\begin{table}[ht]
\centering
\caption{Dataset statistics of RecIF-Bench.}
\label{tab:dataset_stats}
\setlength{\tabcolsep}{1mm}
\renewcommand{\arraystretch}{1.05}
\footnotesize
\begin{tabular}{l rrrrr}
\toprule
\textbf{Domain} & \textbf{Users} & \textbf{Items} & \textbf{Interactions} & \textbf{Avg.\ Len.} & \textbf{Avg.\ Targets} \\
\midrule
Short Video & 156,245 & 11,589,098 & 74,276,520 & 475.38 & 8.93 \\
Ad & 110,858 & 161,305 & 3,613,364 & 32.59 & 5.39 \\
Product & 113,210 & 1,792,738 & 15,352,929 & 135.61 & 6.80 \\
\bottomrule
\end{tabular}
\end{table}

\subsubsection{Baselines.} To evaluate the effectiveness of our method, R$^2$-SFT, we compare against three groups of recommenders. (i) \textit{Traditional discriminative recommenders}: \textbf{SASRec}~\cite{kang2018self}, \textbf{BERT4Rec}~\cite{sun2019bert4rec}, \textbf{GRU4Rec}~\cite{hidasi2015session}, \textbf{HSTU}~\cite{zhai2024actions}, and \textbf{ReaRec}~\cite{tang2026think}. (ii) \textit{Generative recommenders}: \textbf{TIGER}~\cite{rajput2023recommender} and \textbf{LC-Rec}~\cite{zheng2024adapting}. These two groups do not expose a thinking/non-thinking toggle, so we report a single result per domain, taken from~\cite{zhou2025openonerec}. 
(iii) \textit{Foundation recommender models}: \textbf{OpenOneRec-8B}~\cite{zhou2025openonerec}\footnote{https://huggingface.co/OpenOneRec/OneRec-8B}, and \textbf{OpenOneRec-8B-SFT}, which we reproduce by applying the original OpenOneRec SFT recipe on top of the released pretrain checkpoint. Both support the thinking and non-thinking modes, and we re-evaluate them with our own pipeline, which follows the OpenOneRec framework exactly with only minor parameter differences.

\subsubsection{Implementation details.}
Our training and evaluation framework is identical to OpenOneRec\footnote{https://github.com/Kuaishou-OneRec/OpenOneRec}. The base model is the publicly released 8B OpenOneRec pretrain checkpoint\footnote{https://huggingface.co/OpenOneRec/OneRec-8B-pretrain}, built on the Qwen3-8B backbone~\cite{yang2025qwen3}; items are represented as 3-level Semantic IDs, with each level drawn from a codebook of size 8192. The CoT teacher used for rationale synthesis is Gemini~2.5~Flash-Lite. For SID-PPL filtering, we retain approximately $50\%$ of the synthesized CoT samples with the best scores. The warm-up SFT for the PPL scorer runs for roughly $6\%$ of the full unfiltered mixed corpus. The main SFT is trained for 2 epochs with the AdamW optimizer \cite{loshchilov2017decoupled} and a cosine schedule~\cite{loshchilov2016sgdr} (peak learning rate $5\times10^{-6}$, minimum $10^{-7}$, 50 warmup steps, maximum sequence length 32768, gradient clipping at $1.0$). The target loss weight $\lambda$ is set to $1.0$. 

\subsection{Main Results}
\label{subsec:main_results}

Table~\ref{tab:main} reports the overall comparison on the three RecIF-Bench domains. We highlight three observations.

\begin{table*}[t]
\centering
\caption{Overall comparison on RecIF-Bench across three domains. \textbf{Bold} marks the best result per column; \underline{Underlined} values mark the strongest baseline.}
\label{tab:main}
\setlength{\tabcolsep}{1.5mm}
\renewcommand{\arraystretch}{1.05}
\begin{tabular}{lll cc cc cc}
\toprule
\multirow{2}{*}{\textbf{Category}} & \multirow{2}{*}{\textbf{Model}} & \multirow{2}{*}{\textbf{Mode}} & \multicolumn{2}{c}{\textbf{Ad}} & \multicolumn{2}{c}{\textbf{Product}} & \multicolumn{2}{c}{\textbf{Video}} \\
\cmidrule(lr){4-5} \cmidrule(lr){6-7} \cmidrule(lr){8-9}
 & & & Pass@32 & Recall@32 & Pass@32 & Recall@32 & Pass@32 & Recall@32 \\
\midrule
\multirow{5}{*}{\textit{\textbf{Discriminative}}}
 & \textbf{SASRec}   & Non-thinking & 0.0980 & 0.0293 & 0.0914 & 0.0175 & 0.1003 & 0.0119 \\
 & \textbf{BERT4Rec} & Non-thinking & 0.1225 & 0.0381 & 0.0936 & 0.0193 & 0.0951 & 0.0113 \\
 & \textbf{GRU4Rec}  & Non-thinking & 0.1102 & 0.0336 & 0.0821 & 0.0161 & 0.0993 & 0.0117 \\
 & \textbf{HSTU}     & Non-thinking & 0.1266 & 0.0409 & 0.0914 & 0.0178 & 0.1010 & 0.0119 \\
 & \textbf{ReaRec}   & Thinking     & 0.1054 & 0.0327 & 0.0907 & 0.0189 & 0.1002 & 0.0120 \\
\midrule
\multirow{2}{*}{\textit{\textbf{Generative}}}
 & \textbf{TIGER}     & Non-thinking & 0.1769 & 0.0581 & 0.1276 & 0.0283 & 0.1061 & 0.0132 \\
 & \textbf{LC-Rec-8B} & Non-thinking & 0.2096 & 0.0723 & 0.1809 & 0.0416 & 0.1306 & 0.0180 \\
\midrule
\multirow{6}{*}{\textit{\textbf{Foundation}}}
 & \multirow{2}{*}{\textbf{OpenOneRec-8B-SFT}} & Thinking     & 0.1621 & 0.0536 & 0.1148 & 0.0256 & 0.1573 & 0.0267 \\
 &  & Non-thinking & 0.2370 & 0.0828 & 0.1896 & 0.0447 & 0.2042 & 0.0334 \\
 & \multirow{2}{*}{\textbf{OpenOneRec-8B}}     & Thinking     & 0.2036 & 0.0691 & 0.1470 & 0.0337 & 0.1636 & 0.0288 \\
 &  & Non-thinking & \underline{0.2500} & \underline{0.0879} & \underline{0.1964} & \underline{0.0469} & \underline{0.2091} & \underline{0.0354} \\
 & \multirow{2}{*}{\textbf{R$^2$-SFT}}    & Thinking     & \textbf{0.2730} & \textbf{0.0986} & \textbf{0.1980} & \textbf{0.0476} & \textbf{0.2626} & \textbf{0.0422} \\
 &  & Non-thinking & 0.2580 & 0.0919 & 0.1883 & 0.0444 & 0.2378 & 0.0389 \\
\bottomrule
\end{tabular}
\end{table*}

\textbf{(1) R$^2$-SFT substantially improves thinking-mode recommendation over the SFT baseline.} Our training starts from the OpenOneRec-8B pretrain checkpoint, and the directly comparable baseline is OpenOneRec-8B-SFT, which reproduces the original OpenOneRec SFT recipe from the same starting point. Under this recipe, the thinking mode lags far behind the non-thinking mode---on average, thinking-mode Pass@32 trails non-thinking by about $31\%$ across the three domains---confirming that an SFT recipe without effective design for the reasoning path fails to bring gains to the thinking mode. R$^2$-SFT reverses this pattern: unlike OpenOneRec, its thinking mode consistently outperforms its non-thinking mode on all three domains, while also improving Pass@32 by $67$--$72\%$ over OpenOneRec-8B-SFT. This demonstrates that target-blind CoT, SID-level rejection sampling, and the target loss together enable the thinking mode to deliver genuine gains at the SFT stage.

\textbf{(2) The thinking mode now surpasses the strongest baseline.} The strongest baseline overall is the released OpenOneRec-8B in non-thinking mode, which has already been strengthened by distillation and RL on top of the SFT stage. Remarkably, R$^2$-SFT in thinking mode surpasses this stronger checkpoint on Ad and Video ($+9.2\%$ and $+25.6\%$ in Pass@32, respectively) and is on par on Product, even though our method only improves the SFT stage and uses no RL. This indicates that a well-designed SFT alone can unlock the potential of reasoning-style recommendation.

\textbf{(3) Non-thinking performance is preserved and sometimes improved.} A natural concern is whether stronger thinking-mode supervision hurts the non-thinking ability. The results show this is not the case: in non-thinking mode, R$^2$-SFT substantially improves over OpenOneRec-8B-SFT on Ad and Video, and remains on par with it on Product. The on-par result on Product is reasonable, because R$^2$-SFT mainly changes the construction, selection, and optimization of CoT recommendation data, while the non-CoT recommendation data and the standard direct-prediction objective are not specifically modified. Thus, the non-thinking mode is primarily expected to preserve the original direct recommendation ability. The gains on Ad and Video may partly come from the improved thinking ability being internalized into the model's general recommendation capability, thereby benefiting non-thinking inference as well, which is consistent with the observation at the SFT stage in OneReason~\cite{team2026onereason}.

\begin{figure*}[t]
    \centering
    \includegraphics[width=\textwidth]{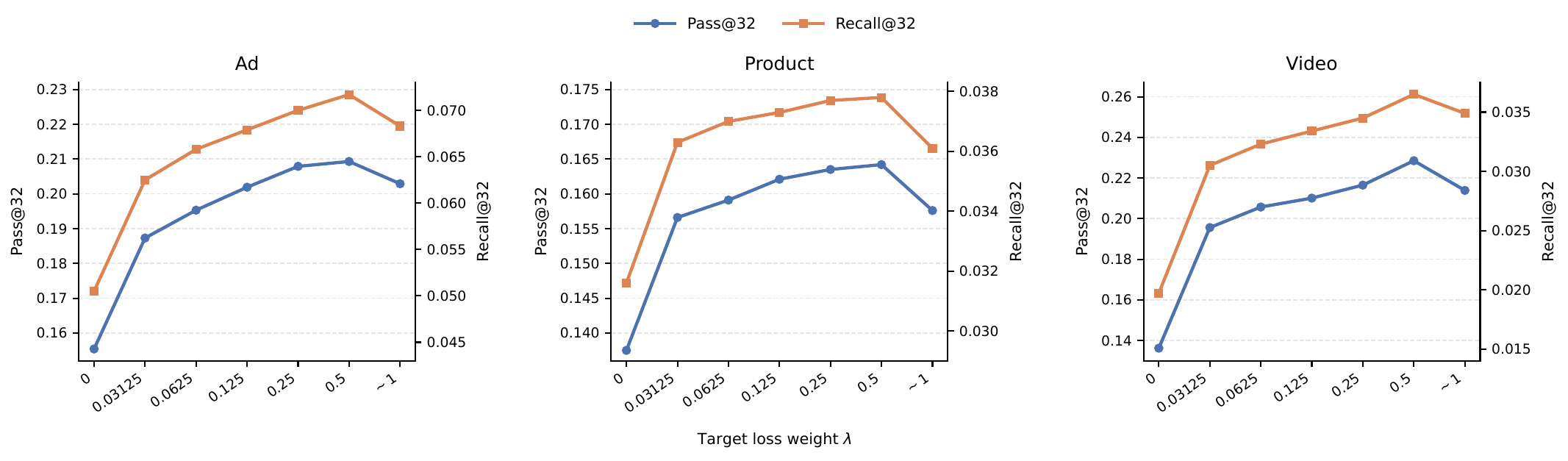}
    \caption{Effect of the target loss weight $\lambda$ in thinking mode. Each column corresponds to one domain, and $\lambda=0.999$ is denoted as $\sim1$. Higher values indicate better recommendation performance.}
    \label{fig:target_loss_sweep}
\end{figure*}

\subsection{Ablation Study}
\label{subsec:ablation}

We conduct a set of controlled ablations to validate the core design choices of R$^2$-SFT. Due to limited computational resources, all ablation experiments are conducted in the thinking mode using only 50\% of the CoT recommendation data, without including the other two types of training data.

\subsubsection{Component Ablation}
\label{subsec:component_ablation}

Table~\ref{tab:ablation} first studies the contribution of the three main components: SID-PPL filtering, the target loss, and target-blind synthesis. The ``Base'' variant uses the complete R$^2$-SFT recipe under the reduced CoT-only setting.

\begin{table}[t]
\centering
\caption{Ablation study on the three core components of R$^2$-SFT (thinking mode). P@32 and R@32 denote Pass@32 and Recall@32, respectively.}
\label{tab:ablation}
\setlength{\tabcolsep}{1mm}
\renewcommand{\arraystretch}{1.05}
\footnotesize
\begin{tabular}{l ccc ccc}
\toprule
\multirow{2}{*}{\textbf{Variant}} & \multicolumn{2}{c}{\textbf{Ad}} & \multicolumn{2}{c}{\textbf{Product}} & \multicolumn{2}{c}{\textbf{Video}} \\
\cmidrule(lr){2-3} \cmidrule(lr){4-5} \cmidrule(lr){6-7}
 & P@32 & R@32 & P@32 & R@32 & P@32 & R@32 \\
\midrule
Base (R$^2$-SFT)             & \textbf{0.2124} & \textbf{0.0718} & \textbf{0.1659} & \textbf{0.0382} & \textbf{0.2276} & \textbf{0.0362} \\
w/o SID-PPL filtering        & 0.2062 & 0.0696 & 0.1550 & 0.0357 & 0.1930 & 0.0301 \\
w/o target loss             & 0.1554 & 0.0505 & 0.1375 & 0.0316 & 0.1362 & 0.0197 \\
w/o target-blind synth.          & 0.1229 & 0.0392 & 0.1076 & 0.0234 & 0.2093 & 0.0333 \\
\bottomrule
\end{tabular}
\end{table}

Table~\ref{tab:ablation} shows that all three components are necessary. Removing SID-PPL filtering consistently hurts performance, especially on Video ($-15.2\%$ Pass@32), indicating that unfiltered target-blind rationales still contain noise. Removing the target loss leads to substantial drops across all domains ($-26.9\%$, $-17.1\%$, and $-40.2\%$ Pass@32 on Ad, Product, and Video), confirming the need to streng-then supervision on the final SID. Replacing target-blind synthesis with target-exposed synthesis causes the largest degradation on Ad and Product, showing that post-hoc rationales do not provide reliable training signals.

\subsubsection{Filtering Strategy Comparison}
\label{subsec:filter_comparison}

\begin{table}[t]
\centering
\caption{Comparison of different CoT filtering strategies (thinking mode).}
\label{tab:filter_compare}
\setlength{\tabcolsep}{1.5mm}
\renewcommand{\arraystretch}{1.05}
\footnotesize
\begin{tabular}{l ccc ccc}
\toprule
\multirow{2}{*}{\textbf{Filtering strategy}} & \multicolumn{2}{c}{\textbf{Ad}} & \multicolumn{2}{c}{\textbf{Product}} & \multicolumn{2}{c}{\textbf{Video}} \\
\cmidrule(lr){2-3} \cmidrule(lr){4-5} \cmidrule(lr){6-7}
 & P@32 & R@32 & P@32 & R@32 & P@32 & R@32 \\
\midrule
No filtering   & 0.2062 & 0.0696 & 0.1550 & 0.0357 & 0.1930 & 0.0301 \\
SID-PPL              & \textbf{0.2124} & \textbf{0.0718} & \textbf{0.1659} & \textbf{0.0382} & \textbf{0.2276} & \textbf{0.0362} \\
Caption-PPL                  & 0.1863 & 0.0621 & 0.1240 & 0.0261 & 0.1914 & 0.0307 \\
Text similarity  & 0.1988 & 0.0663 & 0.1478 & 0.0342 & 0.1684 & 0.0268 \\
LLM-Judge                 & 0.2076 & 0.0708 & 0.1606 & 0.0363 & 0.1745 & 0.0280 \\
\bottomrule

\end{tabular}
\end{table}

We next examine the choice of filtering criterion. All variants are trained with only CoT data and use the same number of training steps, differing only in how synthesized rationales are selected: no filtering, \textbf{SID-PPL} (ours), \textbf{Caption-PPL}, \textbf{text similarity}, or \textbf{LLM-Judge}. 
For the text-level baselines, LLM-Judge assigns a 1--5 score based on the CoT and the target item, while Caption-PPL uses a text model to score the target-caption likelihood under the CoT and input captions. Table~\ref{tab:filter_compare} reports the results.

The SID-PPL filtering strategy gives the best Pass@32 and Recall@32 across the three domains, suggesting that a filtering signal directly tied to SID prediction is more robust in this setting. Text-based criteria, by contrast, are more sensitive to domain-specific noise. Caption-PPL performs particularly poorly on Product, likely because product titles are often short, noisy, or not fully fluent natural-language descriptions, making caption likelihood a weak proxy for recommendation quality. Text similarity and LLM-Judge also do not consistently improve over no filtering, indicating that text-based filtering may need domain-aware tuning.

\subsubsection{Sensitivity to Target Loss Weight}
\label{subsec:target_loss_sweep}

For this sensitivity analysis, we vary the relative weight of the target loss while keeping the overall loss scale fixed. Specifically, we use CE weight $\beta$ and target-loss weight $\lambda$ with $\beta+\lambda=1$, and evaluate seven values of $\lambda$ from $0$ to $0.999$ (shown as $\sim1$ in Figure~\ref{fig:target_loss_sweep}). All runs use the same CoT-only 50\% training setting. 

The performance first increases as $\lambda$ grows from $0$, reaches its best or near-best value around $\lambda=0.5$, and then slightly declines when $\lambda$ approaches $1$. This trend gives two complementary messages. First, $\lambda=0$ removes the additional target supervision and performs poorly, confirming that the target loss is necessary. Second, an overly large $\lambda$ reduces the relative weight of the CE loss and weakens learning on the reasoning trace itself, showing that the original CE objective remains necessary. We therefore use the target loss as a balanced auxiliary objective rather than replacing the standard CE objective.

\subsection{Further Analysis}
\label{subsec:further_analysis}

\subsubsection{Text-level vs.\ SID-level Quality Signals}
\label{subsec:analysis_text_vs_sid}

The filtering analysis in Section~\ref{subsec:filter_comparison} shows that SID-PPL filtering outperforms text-level alternatives, but it does not explain \emph{why} text-level signals fail. We hypothesize that text-level quality signals are largely decoupled from the recommendation objective, because the final prediction is made over SID tokens rather than natural-language captions. 
To verify this, we run two analyses over the same CoT candidates.

\begin{figure}[ht]
    \centering
    \includegraphics[width=\columnwidth]{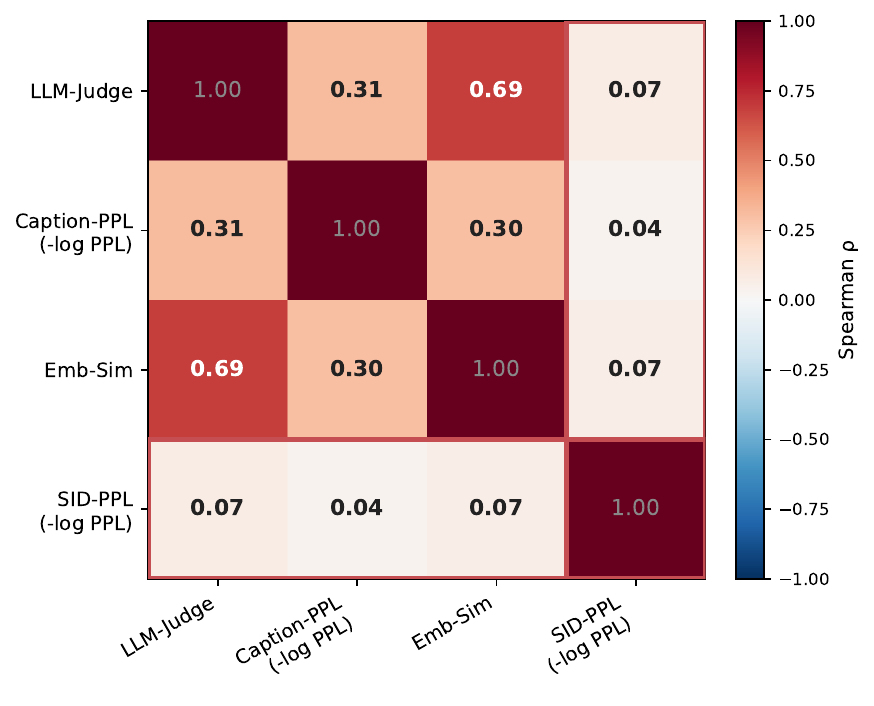}
    \caption{Pairwise Spearman correlation of four CoT quality signals. The three text-level signals (LLM-Judge, Caption-PPL, Emb-Sim) cluster among themselves, but are almost uncorrelated with SID-PPL. This shows that text-level quality is a poor proxy for SID-level recommendation quality.}
    \label{fig:heatmap}
    \vspace{-10pt}
\end{figure}

\paragraph{Pairwise correlation of quality signals.}
For each candidate rationale, we compute four quality signals: the LLM-Judge score, the Caption-PPL score (implemented as the negative log PPL of the target caption given the rationale), the embedding cosine similarity between the rationale and the target caption, and our SID-PPL score (implemented as the negative log PPL of the target SID given the rationale). Figure~\ref{fig:heatmap} shows the pairwise Spearman correlation across the full candidate pool.

The three text-level signals are mutually correlated (LLM-Judge and Emb-Sim reach $\rho=0.69$), indicating that they measure overlapping aspects of textual quality. However, all three are nearly uncorrelated with SID-PPL (mean $\rho=0.07$), which directly measures how much a rationale helps predict the target SID. This means that a rationale can look perfectly fluent and relevant in the caption space while providing little to no evidence for the SID-space prediction, and vice versa. Filtering with a text-level signal therefore selects rationales that are good ``captions'' but not necessarily good ``recommendation rationales.''

\begin{figure*}[t]
    \centering
    \includegraphics[width=0.8\textwidth]{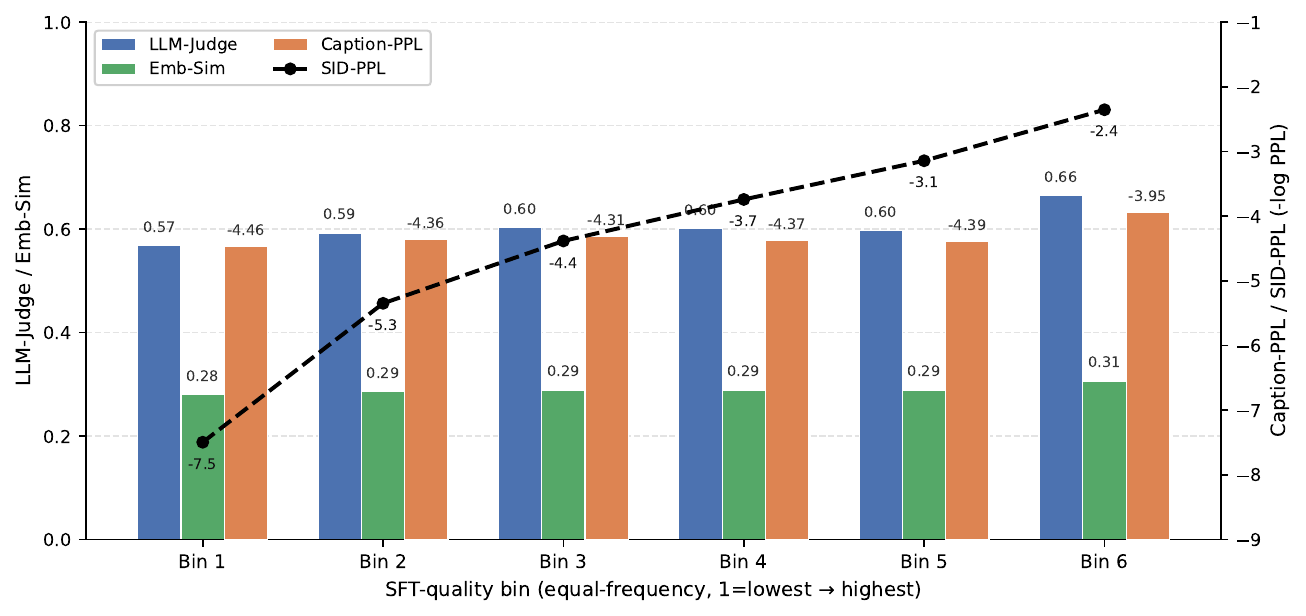}
    \vspace{-5pt}
    \caption{Mean text-level quality scores assigned to candidate rationales after grouping them into six equal-frequency SID-PPL quality bins (1 = lowest, 6 = highest), where SID-PPL is implemented as the negative log PPL of the target SID. Bars show the mean LLM-Judge, embedding-similarity, and Caption-PPL scores, while the dashed black line reports the mean SID-PPL score on the right $y$-axis. The nearly flat bars indicate that these text-level scores vary little as SID-PPL quality improves.}
    \label{fig:binned}
\end{figure*}

\paragraph{Text signals across SID-quality bins.}
To further illustrate the gap, we bin all candidates into six equal-frequency groups by their SID-PPL score, implemented here as the negative log PPL of the target SID (Bin~1 = lowest SID quality, Bin~6 = highest), and compute the mean of each text-level signal within each bin. If a text-level signal tracked recommendation quality, its mean should rise monotonically from Bin~1 to Bin~6. Figure~\ref{fig:binned} shows the result.

The bars are nearly flat across bins for all three text-level signals: LLM-Judge rises only from $0.568$ to $0.665$ ($\Delta=+0.10$), Emb-Sim from $0.280$ to $0.306$ ($\Delta=+0.03$), and negative log Caption-PPL from $-4.46$ to $-3.95$ ($\Delta=+0.52$), whereas the SID-PPL quality score itself rises steeply (dashed line). In other words, the text-level signals barely distinguish the worst SID-quality rationales from the best ones. This provides a direct explanation for the results in Table~\ref{tab:filter_compare}: text-level filtering cannot reliably retain the rationales that actually help recommendation, because the signal it optimizes is decoupled from the SID-space prediction target. SID-PPL, by contrast, directly measures the quantity we care about, making it the only filtering strategy that consistently improves final recommendation accuracy.

\subsubsection{Target-Blind CoT Generalization}
\label{subsec:generalization}

\begin{figure}[t]
    \centering
    \includegraphics[width=0.85\columnwidth]{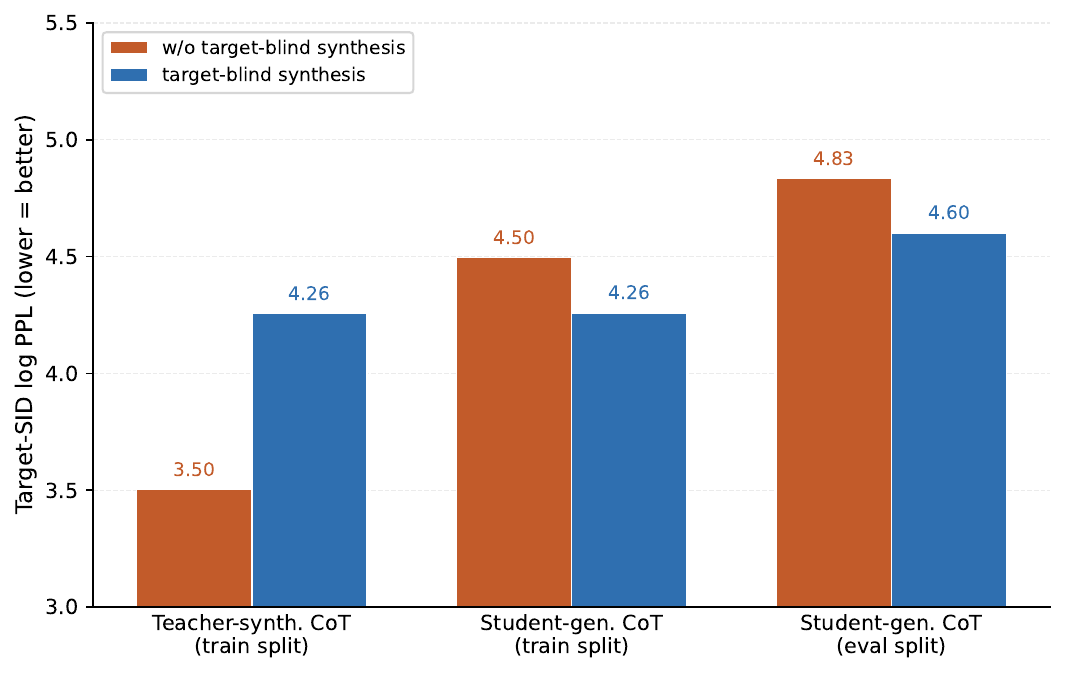}
    \caption{Target-SID log PPL for target-exposed vs. target-blind synthesis. Teacher-synth. denotes rationales synthesized by the teacher model; student-gen. denotes rationales generated by the trained recommendation model. Without target-blind synthesis, the recommendation benefit from teacher-synthesized rationales on the training set transfers poorly to student-generated rationales.}
    \label{fig:generalization}
\end{figure}

The component ablation in Table~\ref{tab:ablation} shows that replacing target-blind synthesis with target-exposed synthesis (``w/o target-blind synth.'') causes a severe drop on Ad and Product. If the synthesized rationale were merely unhelpful, the model could in principle learn to ignore it and behave similarly to the non-thinking mode; the larger degradation suggests that target-exposed synthesis may introduce misleading reasoning patterns. We hypothesize that exposing the target helps the teacher produce rationales that fit the training answers, but such post-hoc rationales are difficult for the student model to reproduce at inference time. To verify this hypothesis, we compare the recommendation quality of target-exposed and target-blind synthesis under three settings: teacher-synthesized rationales on the training set, student-generated rationales on the training set, and student-generated rationales on the evaluation set. For efficiency, we use target-SID log perplexity (PPL) as a proxy for Pass@32. Figure~\ref{fig:generalization} reports the results.

The figure shows a reversal between teacher synthesis and student generation. For teacher-synthesized rationales on the training set, target-exposed synthesis has a lower PPL ($3.50$) than target-blind synthesis ($4.26$): when the teacher is given the target, it can write a rationale that fits the answer, so the target SID appears more predictable. However, after student training, the trend reverses. The student produces higher PPL when trained with rationales from target-exposed synthesis, both on the training set ($4.50$ vs $4.26$) and on the evaluation set ($4.83$ vs $4.60$). This indicates that rationales fitted to known answers are easier for the teacher to synthesize, but harder for the student to use as a transferable reasoning pattern.

\section{Conclusion}
\label{sec:conclusion}

In this work, we studied why explicit reasoning does not automatically improve LLM-based recommendation during supervised fine-tuning stage. Starting from OpenOneRec~\cite{zhou2025openonerec}, we found that the thinking mode can substantially underperform the non-thinking mode, indicating that simply adding CoT supervision is insufficient. We traced this problem to both data and training factors: target-exposed synthesis can produce post-hoc rationales that are hard for the student to generalize, text-level filtering signals are weakly aligned with SID-level recommendation quality, and the standard SFT objective dilutes the supervision on the final target SID.

To address these issues, we proposed R$^2$-SFT for LLM-based generative recommendation. On the data side, R$^2$-SFT uses target-blind CoT synthesis and SID-PPL rejection sampling to obtain rationales that are better aligned with the recommendation objective. On the training side, it introduces an additional target loss to strengthen supervision on the final SID while preserving the original CE objective. Experiments on three industrial domains show that R$^2$-SFT consistently improves thinking-mode recommendation, enables the thinking mode to surpass the non-thinking mode at the SFT stage, and remains competitive with stronger OpenOneRec checkpoints trained with more advanced recipes.

This work suggests that the effectiveness of reasoning in recommendation depends not only on introducing rationales, but also on how they are synthesized, selected, and optimized. One limitation of our study is that the rationale quality is still evaluated mainly through outcome-level signals, such as whether the rationale helps predict the target SID. Future work may develop process-level evaluation methods that directly assess the faithfulness and utility of the intermediate reasoning process, and may further combine R$^2$-SFT with reinforcement learning or preference optimization.


\begin{acks}
This work was supported by the Scientific Research Innovation Capability Support Project for Young Faculty (SRICSPYF-ZY2025008) and the National Natural Science Foundation of China (NSFC) under Grant No. U24A20253.
\end{acks}




\newpage
\bibliographystyle{ACM-Reference-Format}
\bibliography{main}

\end{document}